\documentclass{article} 

\usepackage{preambles}

\numberwithin{equation}{section}
\allowdisplaybreaks

\begin{document}

\hfuzz=100pt
\renewcommand{\thefootnote}{\fnsymbol{footnote}}


\title{
		Comments on the Atiyah-Patodi-Singer index theorem, domain wall, and Berry phase
}
\author{
	\Large
		Tetsuya Onogi $^{1}$ \footnote{onogi(at)phys.sci.osaka-u.ac.jp}, \quad
		Takuya Yoda $^{2}$ \footnote{t.yoda(at)gauge.scphys.kyoto-u.ac.jp}
	\vspace{1em} \\
	\\
	{$^{1}$ \small{\it
		Department of Physics, Osaka University, Toyonaka, Osaka 560-0043, Japan
	}} \\
	{$^{2}$ \small{\it
		Department of Physics, Kyoto University, Kyoto 606-8502, Japan
	}}
}
\date{\small{September 2021}}
\maketitle
\thispagestyle{empty}
\centerline{}


\begin{abstract}
	It is known that the Atiyah-Patodi-Singer index can be reformulated as the eta invariant of the Dirac operators with a domain wall mass which plays a key role in the anomaly inflow of the topological insulator with boundary.
	In this paper, we give a conjecture that the reformulated version of the Atiyah-Patodi-Singer index can be given simply from the Berry phase associated with domain wall Dirac operators when adiabatic approximation is valid. 
	We explicitly confirm this conjecture for a special case in two dimensions where an analytic calculation is possible.
	The Berry phase is divided into the bulk and the boundary contributions,
	each of which gives the bulk integration of the Chern character and the eta-invariant.
\end{abstract}
\vfill
\noindent

OU-HET-1103,
KUNS-2891

\renewcommand{\thefootnote}{\arabic{footnote}}
\setcounter{footnote}{0}

\newpage
\pagenumbering{arabic}
\setcounter{page}{1}
\tableofcontents



\section{Introduction}
\label{sec:intro}

%
%
%
The Atiyah-Patodi-Singer (APS) index theorem \cite{Atiyah:1975jf,Atiyah:1976jg,Atiyah:1980jh} is a generalization of the Atiyah-Singer (AS) index theorem \cite{atiyah1963,10.2307/1970715}.

%
Let $D$ be a massless Dirac operator on an even-dimensional manifold $\mathcal{M}$ without boundary.
The AS index is defined as the difference of the number of two chiral zero modes
$	\ind_{\text{AS}}{D}
	= \dim\ker D_+ - \dim\ker D_-$.
The AS index theorem states that the AS index 
is equal to the instanton number,
\begin{align}
	\label{eq:as_statement}
	\ind_{\text{AS}}{D}
	= \int_{\mathcal{M}}\ch(F).
\end{align}

%
Next,  let  $D$ be also a massless Dirac operator on an even-dimensional manifold $\mathcal{M}$ with a boundary where the APS boundary condition is imposed so that exponentially growing modes are killed.
The APS index is defined similarly by
$	\ind_{\text{APS}}{D}
	= \dim\ker D_+ - \dim\ker D_-$.

The APS index theorem states that the APS index is equal to the sum of the bulk integral of the Chern character and the eta-invariant,
\begin{align}
	\label{eq:aps_statement}
	\ind_{\text{APS}}{D}
	= \int_{\mathcal{M}}\ch(F) + \frac{1}{2}\eta(i\fs{\nabla}), 
\end{align}
where $i\fs{\nabla}$ be its restriction on $\partial \mathcal{M}$.
The eta-invariant $\eta(i\fs{\nabla})$ is the boundary contribution, formally defined by
\begin{align}
	\eta(i\fs{\nabla})
	= \left(
	    \sum_{n} \sgn \lambda_n
	\right)^{\text{reg}}
\end{align}
where $\lambda_n$ is the eigenvalues of $i\fs{\nabla}$ and $"(\cdots)^{\text{reg}}"$ stands for the quantity which is regularized in a proper way such as zeta function regularization.
The eta-invariant is equal to the Chern-Simons term modulo integer.

Recently, non-perturbative generalization of anomaly inflow has been discussed in terms of the eta-invariant\cite{Witten:1999eg,Witten:2015aba,Witten:2019bou}.
It has been applied to high energy physics and condensed matter physics \cite{Witten:2016cio,Tachikawa:2018njr,Hsieh:2018ifc,Garcia-Etxebarria:2018ajm,Wang:2018qoy,Hsieh:2019iba,Davighi:2019rcd,Wan:2020ynf,Hsieh:2020jpj,Davighi:2020uab,Dabholkar:2019nnc}.
The APS index theorem above describes a special case of anomaly inflow\cite{Callan:1984sa}.
See also \cite{Hortacsu:1980kv} for early work on the APS index theorem in physics.

As for applications to topological insulators, the masslessness of Dirac operators in the APS index theorem is puzzling since typical topological insulators are massive inside the bulk.
This puzzle was solved by reformulating the APS index theorem in terms of Dirac operators with a domain wall mass \cite{Fukaya:2017tsq,Fukaya:2019qlf,Fukaya:2020tjk}.
Such a reformulation allows massive fermions inside the bulk and removes the ``non-local'' APS boundary condition.
Thus, it is well suited for actual topological insulators.
See \cite{Vassilevich:2018aqu,Ivanov:2020fsz} for related works.

For the above reasons, it is desired to understand the domain wall APS index theorem at a deeper level.
However, its proof is technically complicated while it is mathematically rigorous.
The best way to deepen understanding is to re-derive the APS index by elementary calculations.

In this paper, we give a conjecture that the APS index can be given simply from the Berry phase associated with domain wall Dirac operators when adiabatic approximation is valid.
We confirm this conjecture by an analytic calculation for a special case in two dimensions.  
The Berry phase is divided into the bulk and the boundary contributions.
Each of them corresponds to the bulk integral of the Chern character and the eta-invariant, respectively.
We intend to report such a new perspective on the domain wall APS index theorem and the eta-invariant in a specific setup.
The more general proof is left for future works.

Our derivation is distinguished from the TKNN formula \cite{Thouless:1982zz}.
The formula computes a Berry phase in the momentum space, while this paper works in the position space.
This paper clarifies how the edge modes contribute to the eta-invariant by examining the contribution to the Berry phase from the spacetime position near the edge.

The paper is organized as follows.
In Sec.~\ref{sec:dw_aps}, we review the reformulation of the APS index theorem in terms of domain wall Dirac operators.
In Sec.~\ref{sec:conj}, we state our conjecture that the APS index is interpreted as a Berry phase.
In Sec.~\ref{sec:berry}, we derive the APS index from the Berry phase associated with a domain wall Dirac operator on a two-dimensional Euclidean torus.
In Sec.~\ref{sec:conc}, we conclude this paper and discuss possible applications of our results.

\section{Domain wall APS index theorem}
\label{sec:dw_aps}

%

Let us consider, for simplicity, a $d(=2n)$-dimensional Euclidean cylinder $\mathcal{M}$ which has boundaries at $x_d=\pm L_2/2$.
While the original APS index theorem \eqref{eq:aps_statement} was formulated in terms of massless Dirac operators, it has recently been reformulated \cite{Fukaya:2017tsq,Fukaya:2019qlf,Fukaya:2020tjk} in terms of domain wall Dirac operators.
They replaced the cylinder with an infinite one but with a domain wall at $x_d=\pm L_2/2$ and introduced a new index, the domain wall APS index, defined by
\begin{align}
	\ind_{\text{DW}}{\: D}
	= \frac{ \eta(D+\bar{\gamma}m_{\text{DW}}) - \eta(D+\bar{\gamma}m_{\text{PV}}) }{2}.
\end{align}
Here, $m_{\text{DW}}, m_{\text{PV}}$ are the domain wall mass and the Pauli-Villars mass respectively
such that
\begin{align}
	&m_{\text{DW}}(x_d) = +\abs{m}\left[ \sgn(x_d+L_2/2)-\sgn(x_d-L_2/2)-1\right], \\
	&m_{\text{PV}}(x_d) = -\abs{m},
\end{align}
where $\abs{m}>0$ is a large mass
\footnote{
	The amplitudes of $m_{\text{DW}}, m_{\text{PV}}$ are not necessarily required to be the same.
	The Pauli-Villars mass is introduced to fix the phase of the trivial vacuum.
}.
They showed that the domain wall APS index is equal to the conventional APS index as
\begin{align}
	\label{eq:dw_aps_statement}
	\ind_{\text{DW}}{D}
	&= \ind_{\text{APS}}D 
	= \int_{\abs{x_d}\leq L_2/2} \ch(F)
	+ \frac{\eta_{\text{L}} - \eta_{\text{R}}}{2}.
\end{align}
Here, $\eta_{\text{L/R}}$ is the eta-invariant defined at $x_d=\mp L_2/2$.

Eta-invariants can be evaluated using, for example, zeta-function regularization.
Especially for the two-dimensional case
\begin{align}
	\mathcal{M}=S_{L_1}^1\times[-L_2/2,L_2/2]
\end{align}
with
\begin{align}
	D = \bar{\gamma}\gamma^{\mu}(\ptl_{\mu}+iA_{\mu}), \quad
	A_1 = -Bx_2 + \frac{2\pi a}{L_1}, \quad
	A_2 = 0,
\end{align}
eta-invariants can be computed explicitly (See App.~\ref{sec:eta_2dim} and, for example, App.~A of \cite{Fukaya:2017tsq}).
The result is
\begin{align}
	\frac{\eta_{\text{L/R}}}{2}
	&= \frac{1}{2}-a_{\text{L/R}}+[a_{\text{L/R}}],
\end{align}
where $a_{\text{L/R}}$ is the holonomy at $x_2=\mp L_2/2$ defined by
$	a_{\text{L/R}}
	= \left. \frac{A_1L_1}{2\pi}\right|_{x_2=\mp L_2/2}$.
\footnote{
In this setup, domain-wall index can be evaluated as $\ind_{\text{DW}}{D}=[a_{\text{L}}]-[a_{\text{R}}]$.
From Stokes theorem, $\frac{1}{2\pi}\int_{\abs{x_2}\leq L_2/2}\dd[2]{x}F_{12}=a_{\text{L}}-a_{\text{R}}$ holds. 
Combining this with the result of the eta invariants, one finds that the APS index theorem holds in the case.
}

\section{APS index as Berry phase}
\label{sec:conj}
In the paper \cite{Witten:2015aba}, the author discussed that the phase of the partition function for a topological insulator is given by
\begin{align}
	Z = \abs{Z} e^{i\pi \cdot \ind_{\text{APS}}{D}}.
\end{align}
Since the APS index can be reformulated in terms of domain wall Dirac operators, it is expected that a similar relation holds for a manifold equipped with domain walls.

\subsection{Our conjecture}
Since the APS index appears as a phase of the partition function,  one may wonder whether 
it can also be derived from canonical formalism. 
For a domain wall fermion with a slowly changing external gauge field and a slowly changing kink mass, where adiabatic approximation is valid, one can expect that the phase of the partition function can be given by the Berry phase since it is the only phase that appears in the partition function under adiabatic approximation. 
Therefore, one is lead to a conjecture:
\begin{eqnarray}
\mbox{APS index} = \mbox{Berry phase of the domain wall system}
\end{eqnarray}
What is non-trivial is that even the eta invariant at the boundary can be included in the Berry phase. 
This could give a new unified view of the APS index.

In the following, let us remind the reader of the fact that the phase of the partition function is given only by the Berry phase when adiabatic approximation is valid for Euclidean theory.
In a quantum mechanical system with Euclidean time $t$,  the Schr\"{o}dinger equation
is given as 
\begin{eqnarray}
0 = (\partial_t + H(t) )|\Psi\rangle, 
\end{eqnarray}
where $H(t)$ is the snapshot Hamiltonian. 
Let as denote $|n(t)\rangle$ and $E_n(t)$ as the $n$-th eigenstate and the $n$-th eigenvalue 
of the snapshot Hamiltonian.
Using a complete set of eigenstates of the snapshot Hamiltonian at time $t$, 
a general state can be written as 
\begin{eqnarray}
|\Psi\rangle = \sum_n a_n(t) |n(t)\rangle.
\end{eqnarray}
Substituting this into the Schr\"{o}dinger equation, one obtains
\begin{eqnarray}
0 = \sum_n (\dot{a}_n | n\rangle + a_n \partial_t |n\rangle + a_n E_n | n\rangle).
\end{eqnarray}
Multiplying $\langle m |$ from the left, one gets
\begin{eqnarray}
0 =\dot{a}_m + \sum_n a_n \langle m| \partial_t |n\rangle + a_m E_ m
\end{eqnarray}
When adiabatic approximation is valid,  the $m$-th eigenstate does not make a transition to a different state under time evolution, therefore
\begin{eqnarray}
\dot{a}_m  = -(\langle m| \partial_t |m\rangle + E_ m) a_m
\end{eqnarray}
holds.
Defining the Berry connection for the m-th state $\mathcal{A}_m$ as $\mathcal{A}_m = -i \langle m | \partial_t |m \rangle$, the equation reads
\begin{eqnarray}
a_m(t) = a_m(0)  \exp\left[-\int_0^t dt^\prime E_m(t^\prime) - i \int_0^t dt^\prime \mathcal{A}_m(t^\prime) \right].
\end{eqnarray}
This means that the phase of the system can be given only by the Berry phase.

\section{Special example in two dimensions}
\label{sec:berry}

In this section, we will consider a two-dimensional torus where topologically trivial/non-trivial phases are jointed with two domain walls in between.
Considering a special gauge configuration where an analytic calculation is possible, 
we will explicitly derive the phase of its partition function
\begin{align}
	\label{eq:desired_result}
	\vartheta = \pi \cdot \ind_{\text{APS}}{D}
\end{align}
from the Berry phase associated with a domain wall Dirac operator 
and confirm our conjecture in the previous section.

\subsection{Phase of partition function}

Let us consider a domain wall fermion $\psi$ on a Euclidean torus,
\begin{align}
	\mathcal{M} = S^1_{L_1} \times S^1_{L_2'} .
\end{align}
We also introduce a Pauli-Villars field (bosonic ghost) $\chi$ with a Pauli-Villars mass.
The Lagrangian is given as
\begin{align}
	\mathscr{L}
		= \bar{\psi} ( i\fs{D}+m_{\text{DW}} )\psi
			+ \bar{\chi} ( i\fs{D}+m_{\text{PV}} )\chi, 
\end{align}
where the domain wall mass and the Pauli-Villars mass are defined as
\begin{align}
	m_{\text{DW}}(x_2)
			&= +\abs{m}\left(
					\tanh{\frac{x_2+L_2/2}{\epsilon}} - \tanh{\frac{x_2-L_2/2}{\epsilon}} -1
			\right)
			\left(= m(x_2)\right), \\
	m_{\text{PV}}(x_2)
			&= -\abs{m}
\end{align}
where $0<L_2<L_2'$ and where $\epsilon$ is a small constant to regulate the jump of the mass.
The domain wall mass divide the Euclidean torus into two regions:
\begin{align}
	\left\{
		\begin{array}{ll}
			\abs{x_2} < L_2/2 & \text{Non-trivial phase} \\
			L_2/2 < \abs{x_2} < L'_2/2 & \text{Trivial phase}
		\end{array}.
	\right.
\end{align}
Regarding the $x_2$-direction as the Euclidean time,
we can write the partition function as
\begin{align}
	Z[A]
	&= \int \mathcal{D}\bar{\psi}\mathcal{D}\psi \mathcal{D}\bar{\chi}\mathcal{D}\chi\:
				\exp\left[
					- \int_{\mathcal{M}} \dd[2]{x}\: \mathscr{L}(A; \psi,\chi)
				\right] \notag\\
	&= \Tr[ (-1)^F\mathcal{P} e^{-\oint \dd{x_2} \hat{h}(x_2)} ].
\end{align}
Here,  $\hat{h}(x_2)$ is the snapshot-Hamiltonian
and where $(-1)^F$ is the fermion parity.
The trace is only over the states in the Dirac sea
since positive energy eigenstates are exponentially suppressed through the Euclidean time development.
We write the Dirac sea as
\begin{align}
	\ket{\Psi^-}
	\propto \sum_{\sigma\in\text{perm.}} (-1)^\sigma \prod_{i} \Psi^-_{\sigma(i)} \ket{0}.
\end{align}
Here, the index $i$ is some quantum number of the negative energy eigenstates.
Let us denote, by $\vartheta_{\text{DW}}$ and $-\vartheta_{\text{PV}}$, the phases acquired through the Euclidean time development of the fermion and the ghost, respectively.
Then, the partition function is expressed as
\begin{align}
	Z[A]
	= \abs{Z[A]} e^{i\vartheta}, \quad
	\vartheta = \vartheta_{\text{DW}}-\vartheta_{\text{PV}}.
\end{align}
Here, the minus sign of $-\vartheta_{\text{PV}}$ reflects that $\chi$ is a ghost.

\subsection{Adiabatic approximation}

We will compute the phase $\vartheta_{\text{DW}}$.
At the end of Sec.~\ref{sec:berry_cr} and Sec.~\ref{sec:berry_blk_bdy}, we include contributions from the ghost.
In the language of the first-quantization,
the fermion satisfies the Dirac equation,
\begin{align}
	D\Psi = 0, \quad
	D = \bar{\gamma} \left( \fs{D} + m \right).
\end{align}
For later convenience, we choose the gamma matrices as
\begin{align}
	\gamma_1 = -\sigma_3, \quad
	\gamma_2 = \sigma_2, \quad
	\bar{\gamma} = -i\gamma_1\gamma_2 = \sigma_1,
\end{align}
and assume that
\begin{align}
	A_{\mu} = A_{\mu}(x_2).
\end{align}
Then, the wave function $\Psi$ satisfies a Schr\"{o}dinger like equation,
\begin{align}
	- \pdv{x_2} \Psi_{p_1}(x_2)
	= h(x_2)\Psi_{p_1}(x_2),
\end{align}
where the snapshot-Hamiltonian $h(x_2)$ is expressed as
\begin{align}
	h(x_2) = h'(x_2) + iA_2(x_2), \quad
	h'(x_2) = (p_1+A_1(x_2))\sigma_1 + m(x_2)\sigma_2.
\end{align}
It has positive/negative energy eigenstates such that
\begin{align}
	h'(x_2)\Psi_{p_1}^{\pm}(x_2)
	= \pm \varepsilon_{p_1}(x_2)\Psi_{p_1}^{\pm}(x_2), \quad
	\varepsilon_{p_1}(x_2) = \sqrt{\tilde{p}_1(x_2)^2+m(x_2)^2},
\end{align}
Here, we wrote $\tilde{p}_1 = p_1 + A_1$.
The snapshot wave function is
\begin{align}
	\Psi^+_{p_1}(x_2)
	&= e^{i\alpha^+}\mqty( \cos\theta/2 \\ e^{i\phi}\sin\theta/2 ), \\
	\Psi^-_{p_1}(x_2)
	&= e^{i\alpha^-}\mqty( \sin\theta/2 \\ e^{i(\pi+\phi)}\cos\theta/2 ),
\end{align}
where we introduced the Bloch sphere coordinates such that
\begin{align}
	h'(x_2)
	= \bm{R}\cdot\bm{\sigma}, \quad
	\bm{R} = \varepsilon \mqty( \sin\theta\cos\phi \\ \sin\theta\sin\phi \\ \cos\theta ),
\end{align}
and where two functions $\alpha^{\pm}=\alpha^{\pm}(x_2)$ correspond to $U(1)$-left/right gauge transformations.

\begin{figure}[t]
	\centering
	\includegraphics[width=75mm]{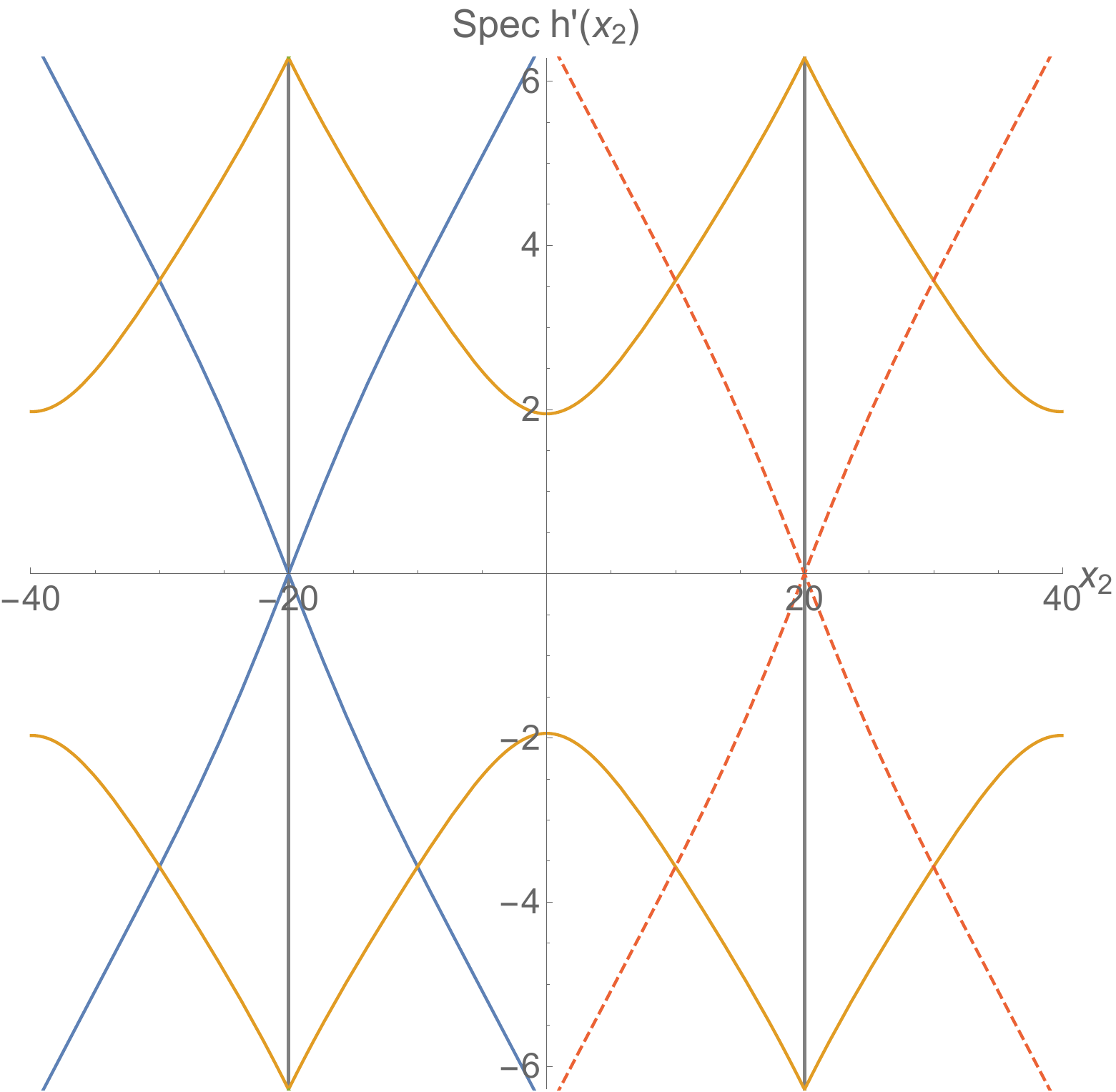} \hspace{5mm}
	\includegraphics[width=75mm]{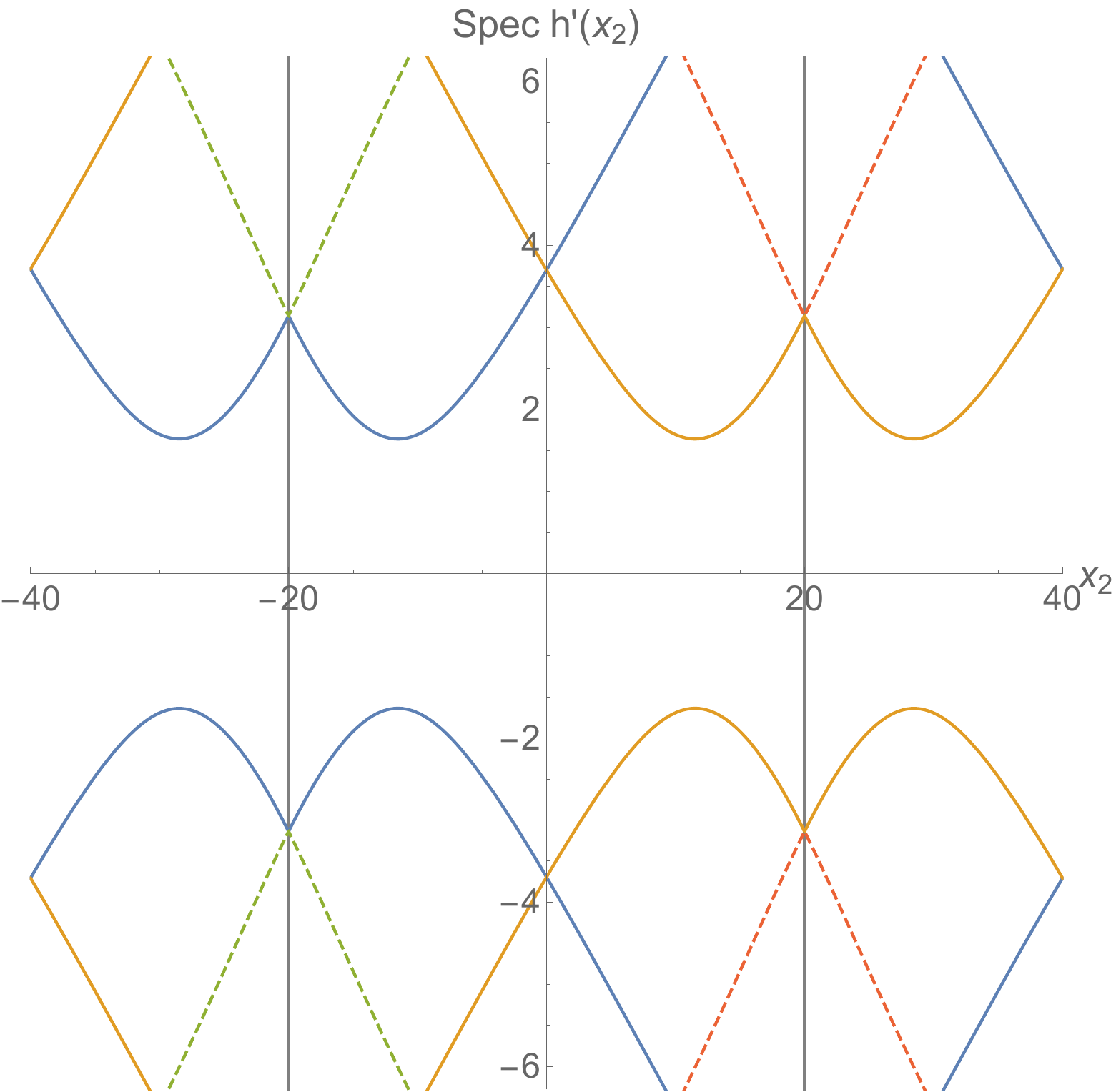}
	\caption{
		Typical spectrum of the snapshot Hamiltonian $h'(x_2)$.
		The green-dashed, blue-solid, orange-solid and red-dashed curves correspond to momentum eigenstates with $p_1L_1/2\pi = -2, -1, 0, 1$, respectively.
		(We will explain the difference of the solid/dashed curves in Sec.~\ref{sec:berry_cr}).
		Domain walls are put on the gray vertical lines.
		The gauge potential monotonically decreases inside the region separated by the domain walls as $A_1(x_2)=-Bx_2+A_1(0) \;(\abs{x_2}<L_2/2)$ and monotonically increases outside as $A_1(x_2)=+B(x_2\mp L_2/2)+A_1(\pm L_2/2) \;(L_2/2<\abs{x_2}<L_2'/2)$, where $B=2\cdot(2\pi/L_1L_2)$.
		Parameters are chosen as $L_1=1, L_2/2=20, \abs{m}=2, \epsilon=8$ so that the hierarchy \eqref{eq:hierarchy} is satisfied.
		In the left panel, the holonomy $a=A_1(0)L_1/2\pi$ is an integer, invalidating adiabatic approximation at the domain walls.
		In the right panel, the holonomy is shifted to a half-integer, validating adiabatic approximation.
	}
	\label{fig:spec}
\end{figure}

A typical spectrum of the snapshot Hamiltonian is shown in Fig.\;\ref{fig:spec}.
%
%
In the following, we assume the following hierarchy of parameters
\begin{align}
		\label{eq:hierarchy}
		\frac{\abs{m}}{\epsilon} \ll L_1^{-2} \ll \abs{m}^2.
\end{align}
Since each momentum eigenstates are decoupled and the bulk mass is sufficiently large
\begin{align}
	\abs{m}L_1 \gg 1
\end{align}
under the hierarchy, the Landau-Zener effect \cite{10011873546,Zener} (See also App.~\ref{sec:lz}) is well suppressed at almost all Euclidean times.
The dangerous time is when the domain wall mass goes to zero $m(x_2)=0$.
However, unless the holonomy $a$ is an integer, the hierarchy assures that the Landau-Zener effect is well suppressed since
\begin{align}
	\frac{\varepsilon_{p_1}^2}{\abs{m}/\epsilon}
	\sim \frac{L_1^{-2}}{\abs{m}/\epsilon}
	\gg 1.
\end{align}
Here, we used \eqref{eq:lz_prob} by replacement $\varepsilon_{12} \rightarrow \varepsilon_{p_1}, \alpha \rightarrow \abs{m}/\epsilon$.
Thus, the adiabatic approximation is valid.
Also, let us assume that $L_2$ is sufficiently large so that the bulk and the boundary regions are well distinguished while the mass gap opens slowly $\epsilon\abs{m}\gg 1$ around the domain walls.
The largeness of $L_2$ is consistent with the previous discussion that the Dirac sea states dominate the functional trace.


%
Under adiabatic approximation, each state acquires a Berry phase,
\begin{align}
	i\ev{-i\ptl_2}{\Psi_{p_1}^-} + iA_2
	&= i\ev{(-i\ptl_2+A_2)}{\Psi_{p_1}^-}.
\end{align}
The first term of the left-hand side is associated with $h'(x_2)$.
Each momentum eigenstate develops as
\begin{align}
	\Psi^-_{p_1}(0)
	\rightarrow \exp\left[
								- \int_0^{x_2}\dd{x_2} \left(
									-\varepsilon_{p_1} + i\ev{(-i\ptl_2+A_2)}{\Psi_{p_1}^-}
								\right)
							\right]
							\Psi^-_{p_1}(0).
\end{align}
Summing over all the states in the Dirac sea, we find that
\begin{align}
	\vartheta_{\text{DW}}
	\simeq \sum_{p_1} \oint \dd{x_2} \ev{(-i\ptl_2+A_2)}{\Psi_{p_1}^-}
\end{align}
as long as the adiabatic approximation is valid.
Note that the phase $\vartheta_{\text{DW}}$ is gauge invariant, reflecting the compactness of the manifold $\mathcal{M}$.
We fix the phase of the wave function so that $\vartheta_{\text{DW}} = 0$ for $A=0$.
It means that we have to set
\begin{align}
	\ptl_2 \alpha^- + A_2 = 0.
\end{align}
In the following computations, we set $\alpha^-(x_2)=A_2(x_2) = 0$ for simplicity.
The total Berry phase becomes
\begin{align}
	\vartheta_{\text{DW}}
	\label{eq:berry_cr}
	&\simeq \frac{1}{2}\sum_{p_1} \oint \dd{x_2}
				\ptl_2\phi (1+\cos\theta)
	= \frac{1}{2}\sum_{p_1} \oint \dd{x_2}
				\ptl_2\phi \\
	\label{eq:berry_blk-bdy}
	&= \frac{1}{2}\sum_{p_1} \oint  \dd{x_2}
	\frac{\tilde{p}_1m}{\tilde{p}_1^2+m^2}
	\left(
	-\frac{\ptl_2\tilde{p}_1}{\tilde{p}_1} + \frac{\ptl_2m}{m}
	\right).
\end{align}
We will evaluate this quantity in two different ways in the following Sec.~\ref{sec:berry_cr} and Sec.~\ref{sec:berry_blk_bdy}.
They will clarify the meanings of the Berry phase from two different perspectives.

\subsection{Berry phase and level crossings}
\label{sec:berry_cr}

Firstly, we evaluate the expression \eqref{eq:berry_cr}.
The following computations will clarify the relationship between our formulation and the level-crossings in the context of the original APS index theorem.

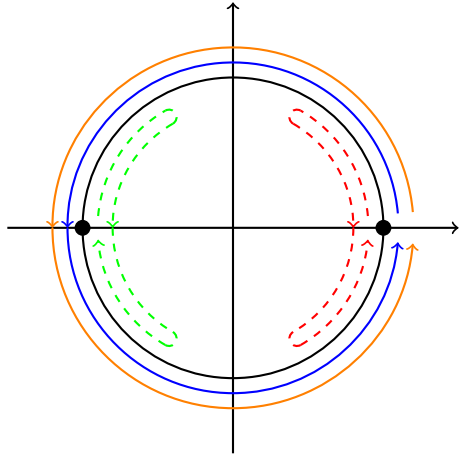
\begin{figure}
	\centering
	\begin{tikzpicture}
		\draw [thick] (0,0) circle (2);
		\draw [->,thick] (-3,0) -- (3,0);
		\draw [->,thick] (0,-3) -- (0,3);
		
		\fill (2,0) circle (3pt);
		\fill (-2,0) circle (3pt);
		
		\draw [->,blue,thick] ([shift={(0,0)}]5:2.2) arc [radius=2.2, start angle=5, end angle= 180];
		\draw [->,blue,thick] ([shift={(0,0)}]180:2.2) arc [radius=2.2, start angle=180, end angle= 355];

		\draw [->,orange,thick] ([shift={(0,0)}]5:2.4) arc [radius=2.4, start angle=5, end angle= 180];
		\draw [->,orange,thick] ([shift={(0,0)}]180:2.4) arc [radius=2.4, start angle=180, end angle= 355];
		
		\draw [red,thick,dashed] ([shift={(0,0)}]5:1.8) arc [radius=1.8, start angle=5, end angle= 60];
		\draw [->,red,thick,dashed] ([shift={(0,0)}]60:1.6) arc [radius=1.6, start angle=60, end angle= 0];
		\draw [red,thick,dashed] ([shift={(0,0)}]0:1.6) arc [radius=1.6, start angle=0, end angle= -60];
		\draw [->,red,thick,dashed] ([shift={(0,0)}]-60:1.8) arc [radius=1.8, start angle=-60, end angle= -5];
		\draw [red,thick,dashed] ([shift={({1.7*cos(60)},{1.7*sin(60)})}]60:0.1) arc [radius=0.1, start angle=60, end angle= 240];
		\draw [red,thick,dashed] ([shift={({1.7*cos(-60)},{1.7*sin(-60)})}]-60:0.1) arc [radius=0.1, start angle=-60, end angle= -240];
		
		\draw [green,thick,dashed] ([shift={(0,0)}]175:1.8) arc [radius=1.8, start angle=175, end angle= 120];
		\draw [->,green,thick,dashed] ([shift={(0,0)}]120:1.6) arc [radius=1.6, start angle=120, end angle= 180];
		\draw [green,thick,dashed] ([shift={(0,0)}]180:1.6) arc [radius=1.6, start angle=180, end angle= 240];
		\draw [->,green,thick,dashed] ([shift={(0,0)}]240:1.8) arc [radius=1.8, start angle=240, end angle=185];
		\draw [green,thick,dashed] ([shift={({1.7*cos(120)},{1.7*sin(120)})}]-60:0.1) arc [radius=0.1, start angle=-60, end angle= 120];
		\draw [green,thick,dashed] ([shift={({1.7*cos(240)},{1.7*sin(240)})}]-120:0.1) arc [radius=0.1, start angle=-120, end angle= 60];
	\end{tikzpicture}
	\caption{
		Typical Euclidean time development along the equator $\theta=\pi/2$ of the Bloch sphere.
		The blue-solid and orange-solid curves are homotopically non-trivial modes, while the green-dashed and red-dashed curves are homotopically trivial modes.
		Each type of the solid/dashed curves corresponds to the ones in Fig.~\ref{fig:spec}.
		The two blobs correspond to the two domain walls.
	}
	\label{fig:homotopy}
\end{figure}

Recall that the Bloch sphere coordinates are related with the Euclidean time as
\begin{align}
	\label{eq:angle}
	\cos\phi = \frac{\tilde{p}_1(x_2)}{\sqrt{\tilde{p}_1(x_2)^2+m(x_2)^2}}, \quad
	\sin\phi = \frac{m(x_2)}{\sqrt{\tilde{p}_1(x_2)^2+m(x_2)^2}}.
\end{align}
Note that the domain wall mass crosses zero $m(x_2)=0$ twice if we go around the torus along the Euclidean time.
Also, note that some momentum eigenstates cross $\tilde{p}_1(x_2)=0$ twice.
For example, the two states illustrated as blue/orange solid curves in the right panel of Fig.~\ref{fig:spec}
cross  $\tilde{p}_1(x_2)=0$ once within $\abs{x_2}<L_2/2$ and within $L_2/2<\abs{x_2}<L_2'/2$.
Then, from $\eqref{eq:angle}$, we find that such states acquire a phase $\Delta\phi=2\pi$ if we go around the torus.
Their wave functions develop as
\begin{align}
	\begin{array}{lccc}
		x_2: & -L_2/2+0 & \rightarrow & -L_2/2-0 \\
		\Psi^-:  & \frac{1}{\sqrt{2}}\mqty( 1\\e^{i\pi} ) & \rightarrow & \frac{1}{\sqrt{2}}\mqty( 1\\e^{i\pi+i2\pi} )
	\end{array}.
\end{align}
They are depicted as the blue/orange (homotopically non-trivial) curves in Fig.\;\ref{fig:homotopy}.

On the other hand,
note that the other momentum eigenstates do not cross $\tilde{p}_1(x_2)=0$.
For example, the two states illustrated as green/red dashed curves in the right panel of Fig.~\ref{fig:spec} do not cross $\tilde{p}_1(x_2)=0$.
Then, from $\eqref{eq:angle}$, we find that such states acquire no phase $\Delta\phi=0$ if we go around the torus.
Their wave functions develop as
\begin{align}
	\begin{array}{lccc}
		x_2: & -L_2/2+0 & \rightarrow & -L_2/2-0 \\
		\Psi^-: & \frac{1}{\sqrt{2}}\mqty( 1\\e^{i\pi}) & \rightarrow & \frac{1}{\sqrt{2}}\mqty( 1\\e^{i\pi} )
	\end{array},
\end{align}
or
\begin{align}
	\begin{array}{lccc}
		x_2: & -L_2/2+0 & \rightarrow & -L_2/2-0 \\
		\Psi^-: & \frac{1}{\sqrt{2}}\mqty( 1\\e^{i\pi+i\pi} ) & \rightarrow & \frac{1}{\sqrt{2}}\mqty( 1\\e^{i\pi+i\pi} )
	\end{array}.
\end{align}
They are depicted as the green/red dashed (homotopically trivial) curves in Fig.\;\ref{fig:homotopy}.

%
%
%
Each of the homotopically non-trivial modes gives an equal contribution $2\pi$ to the Berry phase $\vartheta_{\text{DW}}$.
The total number of  such homotopically non-trivial modes is equal to the number of crossings through $\tilde{p}_1(x_2)=0$.
That is
\begin{align}
	[a_{\text{L}}] - [a_{\text{R}}].
\end{align}
On the other hand, none of the homotopically trivial modes gives any contribution.
Thus, the total Berry phase from the domain wall fermion is
\begin{align}
	\vartheta_{\text{DW}}
	&= \frac{1}{2}\sum_{p_1} \oint \dd{x_2}
				\ptl_2\phi \notag\\
	&= \pi \left(  [a_{\text{L}}] - [a_{\text{R}}] \right) \notag\\
	&= \pi \cdot \ind_{\text{APS}}{D}.
\end{align}

Next, let us consider the ghost contribution.
The Pauli-Villars mass never crosses zero through the Euclidean time development.
Thus, all momentum modes are homotopically trivial even if they cross $\tilde{p}_1(x_2)=0$.
Thus,
\begin{align}
    \vartheta_{\text{PV}}
    =0.
\end{align}

Combining the above results, we obtain the desired result \eqref{eq:desired_result}
\begin{align}
	\label{eq:berry_cr_res}
    \vartheta
    &= \vartheta_{\text{DW}} - \vartheta_{\text{PV}} \notag\\
    &= \pi \cdot \ind_{\text{APS}}{D}.
\end{align}
This result implies that our formulation counts the number of the level crossings as the original APS index theorem does.
However, it appears puzzling that the Pauli-Villars field does not give any contribution as opposed to the usual wisdom.  
\footnote{
For example, a standard calculation shows that the phase of the partition function for a three-dimensional manifold without boundary, which is given by the Chern-Simons action,  one obtains an equal amount of contributions from physical field and Pauli-Villars field. 
}
Actually, this is just a superficial puzzle caused by adding the two regions $\abs{x_2}<L_2/2$ and $L_2/2<\abs{x_2}<L_2'/2$.
In the next Sec.~\ref{sec:berry_blk_bdy}, we will divide the torus into these two regions and see that the Berry phase is ``doubled'' by including the fermion and the ghost contributions.

\subsection{Berry phase and bulk/boundary contributions}
\label{sec:berry_blk_bdy}

Secondly, we evaluate the expression \eqref{eq:berry_blk-bdy}.
The following computations will clarify how the bulk and the boundary modes contribute to the bulk integral of the Chern-character and the eta-invariant respectively.

Let us decompose $\vartheta_{\text{DW}}$ into the inside region $\abs{x_2} < L_2/2$ and the outside region $L_2/2 < \abs{x_2} < L_2'/2$ divided by the domain walls as
\begin{align}
	\vartheta_{\text{DW}}
	= \vartheta_{\text{DW}}^{\text{inside}} + \vartheta_{\text{DW}}^{\text{outside}}.
\end{align}
We focus on the inside part $\vartheta_{\text{DW}}^{\text{inside}}$.
The outside part $\vartheta_{\text{DW}}^{\text{outside}}$ can be computed similarly as the inside one $\vartheta_{\text{DW}}^{\text{inside}}$.
Note that the first term of \eqref{eq:berry_blk-bdy} dominates in the bulk region where $\ptl_2\tilde{p}_1/\ptl_2m \gg 1$ while the second term dominates around the boundary regions where $\ptl_2\tilde{p}_1/\ptl_2m \ll 1$.
We can decompose $\vartheta_{\text{DW}}^{\text{inside}}$ as
\begin{align}
	\vartheta_{\text{DW}}^{\text{inside}}
	\simeq
	\vartheta_{\text{blk}} + \vartheta_{\text{L-bdy}} + \vartheta_{\text{R-bdy}},
\end{align}
where
\begin{align}
	&\vartheta_{\text{blk}}
	= \frac{1}{2}\sum_{p_1} \int_{\text{blk}} \dd{x_2}
	\frac{\abs{m}}{\tilde{p}_1(x_2)^2+\abs{m}^2}(-\ptl_2\tilde{p}_1(x_2)), \\
	&\vartheta_{\text{L-bdy}}
	= \frac{1}{2}\sum_{p_1} \int_{\text{L-bdy}} \dd{x_2}
	\frac{\tilde{p}_1(-L_2/2)}{\tilde{p}_1(-L_2/2)^2+m(x_2)^2} \ptl_2m(x_2), \\
	&\vartheta_{\text{R-bdy}}
	= \frac{1}{2}\sum_{p_1} \int_{\text{R-bdy}} \dd{x_2}
	\frac{\tilde{p}_1(+L_2/2)}{\tilde{p}_1(+L_2/2)^2+m(x_2)^2} \ptl_2m(x_2).
\end{align}
Such a decomposition is justified for the limit $\abs{m}\rightarrow\infty, \epsilon\rightarrow0, L_2\rightarrow\infty$, keeping the hierarchy \eqref{eq:hierarchy}.

\subsubsection{Bulk contribution}

Let us introduce dimensionless quantities
\begin{align}
	&a(x_2) = \frac{A_1(x_2)L_1}{2\pi}, \quad
	b(x_2) = \frac{m(x_2)L_1}{2\pi},
\end{align}
for later convenience.
The bulk contribution is rewritten as
\begin{align}
	\vartheta_{\text{blk}}
	&= \frac{1}{2}\sum_{p_1}\int_{\text{blk}} \dd{x_2}
	\frac{1/\abs{m}}{1+\left(\tilde{p}_1(x_2)/\abs{m}\right)^2}
	(-\ptl_2A_1(x_2)).
\end{align}
Here, note that
\begin{align}
	\frac{\tilde{p}_1(x_2)}{\abs{m}}
	= \frac{2\pi n}{\abs{m}L_1} + \frac{A_1(x_2)}{\abs{m}}
	= \frac{n+a}{\abs{b}}.
\end{align}
Inside the bulk, the mass gap is sufficiently larger than the Compton scale
and than the gauge potential as
\begin{align}
	\abs{b} \gg 1, \quad
	\frac{a}{\abs{b}} \ll 1.
\end{align}
Thus, the summation over the momentum $p_1$ is well approximated by an integral.
Performing the momentum integral first
\begin{align}
	\vartheta_{\text{blk}}
	&= \frac{1}{2}\int_{\text{blk}} \dd[2]{x}
	\int_{-\infty}^{\infty}\frac{\dd{p_1}}{2\pi}
	\frac{1/\abs{m}}{1+\left(\tilde{p}_1(x_2)/\abs{m}\right)^2}
	(-\ptl_2A_1(x_2)) \notag\\
	&= \frac{1}{2}\int_{\text{blk}} \dd[2]{x}
	\int_{-\infty}^{\infty}\frac{\dd{(\tilde{p}_1/\abs{m})}}{2\pi}
	\frac{1}{1+\left(\tilde{p}_1/\abs{m}\right)^2}
	(-\ptl_2A_1(x_2)) \notag\\
	&= \frac{1}{4}\int_{\text{blk}} \dd[2]{x}
	(-\ptl_2A_1(x_2)).
\end{align}
This is nothing but the bulk integral of the Chern-character
\begin{align}
	\vartheta_{\text{blk}}
	= \frac{\pi}{2} \cdot \frac{1}{2\pi} \int \dd[2]{x} F_{12}.
\end{align}

\subsubsection{Boundary contribution}

The left-boundary contribution is rewritten as
\begin{align}
	\vartheta_{\text{L-bdy}}
	&= \frac{1}{2}\sum_{p_1} \int_{\text{L-bdy}} \dd{x_2}
	\frac{\tilde{p}_1(-L_2/2)}{\tilde{p}_1(-L_2/2)^2+m(x_2)^2}\ptl_2m(x_2) \notag\\
	&= \frac{1}{2}\sum_{p_1} \int_{\text{L-bdy}} \dd{m}
	\frac{\tilde{p}_1(-L_2/2)}{\tilde{p}_1(-L_2/2)^2+m^2}.
\end{align}
Here, note that
\begin{align}
	\tilde{p}_1(-L_2/2)
	= \frac{2\pi n}{L_1} + A_1(-L_2/2)
	= \frac{2\pi}{L_1}(n+a_\text{L}).
\end{align}
Recall that we assumed that the compactification length $L_1$ is small so that the adiabatic approximation is valid.
Also, the mass gap is almost zero around the boundary.
Thus, the summation over the momentum $p_1$ is no longer approximated by an integral.
We will perform the integral over $m$ first instead of the summation over the momentum.
Writing the integral
\footnote{
	Another integration interval $-\abs{b}\leq b<0$ belongs to $\vartheta_{\text{DW}}^{\text{outside}}$.
}
with dimensionless quantities,
\begin{align}
	\vartheta_{\text{L-bdy}}
	&= \frac{1}{2}\lim_{\abs{b}\rightarrow\infty}
	\sum_{n} \int_{0}^{\abs{b}} \dd{b}
	\frac{n+a_\text{L}}{(n+a_\text{L})^2+b^2}.
\end{align}
At this stage, we formally find that $\vartheta_{\text{L-bdy}}$ is proportional to the eta-invariant as
\begin{align}
	\vartheta_{\text{L-bdy}}
	&= \frac{1}{2}\sum_n \left[
				\tan^{-1}\frac{b}{n+a_\text{L}}
			\right]^{\infty}_{0} \notag\\
	&= \frac{\pi}{2}\cdot \frac{1}{2} \left[
				\sum_{n+a_\text{L}>0}1 + \sum_{n+a_\text{L}<0}(-1)
			\right] \notag\\
	&= \frac{\pi}{2} \cdot \frac{\eta_{\text{L}}}{2}.
\end{align}
The last equality is from \eqref{eq:eta_form}.
Also, we can derive a regularized version of the eta-invariant.
We decompose the integrand as
\begin{align}
	\label{eq:phase_expr1}
	\vartheta_{\text{L-bdy}}
	&= \frac{1}{2}\lim_{\abs{b}\rightarrow\infty}
	\sum_{n} \int_{0}^{\abs{b}} \dd{b}
	\frac{n+a_\text{L}}{(n+a_\text{L})^2+b^2} \notag\\
	&= \lim_{\abs{b}\rightarrow\infty}
	\frac{1}{4} \sum_{n} \int_{0}^{\abs{b}} \dd{b}
	\left[
	\frac{1}{n+a_\text{L}-ib} + \frac{1}{n+a_\text{L}+ib}
	\right].
\end{align}
Performing the integral over $b$,
\begin{align}
	\vartheta_{\text{L-bdy}}
	&= \lim_{\abs{b}\rightarrow\infty}
	\frac{i}{4} \sum_{n}
	\ln \frac{n+a_\text{L}-i\abs{b}}{n+a_\text{L}+i\abs{b}} \notag\\
	&= \lim_{\abs{b}\rightarrow\infty}
	\frac{i}{4} \ln
	\frac{\sin\pi(a_\text{L}-i\abs{b})}{\sin\pi(a_\text{L}+i\abs{b})}
	= \lim_{\abs{b}\rightarrow\infty}
	\frac{i}{4} \ln
	\frac{ e^{+\pi \abs{b}}e^{+i\pi a_\text{L}} - e^{-\pi \abs{b}}e^{-i\pi a_\text{L}} }
	{ e^{-\pi \abs{b}}e^{+i\pi a_\text{L}} - e^{+\pi \abs{b}}e^{-i\pi a_\text{L}} } \notag\\
	&= \frac{i}{4} \ln e^{2\pi i (a_\text{L} - 1/2)}
	= \frac{\pi}{2} \left( \frac{1}{2} - a_\text{L} + c \right).
\end{align}
At the final line, there is an ambiguity of the choice of the branch $c\in\mathbb{Z}$.
To fix the branch, recall the original expression \eqref{eq:phase_expr1}.
Note that a shift $a_\text{L} \rightarrow a_\text{L}+\mathbb{Z}$ can be absorbed by the shift of the momentum label $n$.
This means that the final result depends on $a_{\text{L}}-[a_{\text{L}}]$ rather than $a_{\text{L}}$ itself.
Also, note that the phase is zero for $a_{\text{L}}=1/2$ due to the symmetry for exchanging $n\leftrightarrow-n-1$.
The only choice to satisfy these conditions is $c=[a_{\text{L}}]$.

Then, we obtain the regularized expression (See \eqref{eq:eta_reg}) for the eta-invariant
\begin{align}
	\vartheta_{\text{L-bdy}}
	= \frac{\pi}{2} \cdot \frac{\eta_\text{L}}{2}.
\end{align}
Similarly, we obtain
\begin{align}
	\vartheta_{\text{R-bdy}}
	= -\frac{\pi}{2} \cdot \frac{\eta_\text{R}}{2}.
\end{align}
The minus sign appears since the mass decreases around the right boundary, contrary to the left boundary.

\subsubsection{Total phase}

Summing over the above results
\begin{align}
	\vartheta_{\text{DW}}^{\text{inside}}
	\simeq \frac{\pi}{2}
	\left[
	\frac{1}{2\pi}\int\dd[2]{x}F_{12}
	+\frac{\eta_\text{L}-\eta_\text{R}}{2}
	\right]
	= \frac{\pi}{2} \cdot \ind_{\text{APS}}{D}.
\end{align}
This is half of the desired result \eqref{eq:desired_result}.
However, including the outside region and the ghost contributions, we finally arrive at
\begin{align}
	\vartheta
	&= \vartheta_{\text{DW}} - \vartheta_{\text{PV}} \notag\\
	&= \left[ \vartheta_{\text{DW}}^{\text{inside}}+\vartheta_{\text{DW}}^{\text{outside}} \right]
			- \left[ \vartheta_{\text{PV}}^{\text{inside}}+\vartheta_{\text{PV}}^{\text{outside}} \right] \notag\\
	&= \frac{\pi}{2}\left[
				\left( +\frac{1}{2\pi}\int_{\text{inside}}\dd[2]{x}F_{12} + \frac{\eta_\text{L}-\eta_\text{R}}{2} \right)
				+\left( -\frac{1}{2\pi}\int_{\text{outside}}\dd[2]{x}F_{12} + \frac{\eta_\text{L}-\eta_\text{R}}{2} \right)
			\right] \notag\\
			&\hspace{10mm}
			-\frac{\pi}{2}\left[
				\left( -\frac{1}{2\pi}\int_{\text{inside}}\dd[2]{x}F_{12}+0 \right)
				+\left( -\frac{1}{2\pi}\int_{\text{outside}}\dd[2]{x}F_{12} + 0 \right)
			\right] \notag\\
	&= \pi \left[
				\frac{1}{2\pi}\int_{\text{inside}}\dd[2]{x}F_{12} + \frac{\eta_\text{L}-\eta_\text{R}}{2}
			\right]
	= \pi \cdot \ind_{\text{APS}}{D}.
\end{align}
Here, the sign in front of $\int F_{12}$ reflects the sign of the mass in the inside/outside region.
Note that the bulk integral of Chern-character is doubled due to the Pauli-Villars contribution, while the eta-invariants are doubled due to contributions from another side of the domain walls.
The outside bulk contributions cancel, implying that the Pauli-Villars term nicely chose the phase of the trivial vacuum.

Now, we can answer the puzzle at the end of Sec.~\ref{sec:berry_cr}.
Due to the periodicity of the gauge potential on the torus, $\int F_{12}$ over the inside/outside region changes its sign.
Thus,
\begin{align}
	\vartheta_{\text{DW}}
	&= 2\cdot \frac{\pi}{2}\left(
			+\frac{1}{2\pi} \int_{\text{inside}} \dd[2]{x} F_{12}
			+\frac{\eta_{\text{L}}-\eta_{\text{R}}}{2}
		\right) \notag\\
	&= \pi\cdot \ind_{\text{APS}}{D}, \\
	\vartheta_{\text{PV}}
	&= 0.
\end{align}
These agree with the previous result \eqref{eq:berry_cr_res}.

Finally, let us mention the case where $\frac{1}{2\pi}\int_{T^2} F_{12}$ takes a non-zero integer value $m$.
In this case, the Dirac sea state does not come back to the original state as we go around the torus.
Rather, its momentum eigenstates shift as $\Psi^{-}_{n}\rightarrow\Psi^{-}_{n-m}$.
Thus, the Dirac sea state acquire a phase as
\begin{align}
	\Psi^{-}_{n_1}\Psi^{-}_{n_2}\cdots\ket{0}
	&\rightarrow
	e^{i\theta_{\text{DW}}^{(1)}}\Psi^{-}_{n_1-m}\:
	e^{i\theta_{\text{DW}}^{(2)}}\Psi^{-}_{n_2-m}\cdots\ket{0} \notag\\
	&= (-1)^{f(m)}e^{i\theta_{\text{DW}}}
	\Psi^{-}_{n_1}\Psi^{-}_{n_2}\cdots\ket{0},
\end{align}
where $f(m)$ is some function obtained by computing infinite products with an appropriate regularization.
If we normalize the partition function so that its phase is zero for $A_{\mu}=0$ as we did in this whole paper, the prefactor $(-1)^{f(m)}$ is simply dropped.
A similar discussion holds for the ghost with the Pauli-Villars mass.
Thus, our conjecture that the APS index is equal to the Berry phase is unchanged.


\section{Conclusion and Discussions}
\label{sec:conc}

In this paper, we  proposed  that the APS index can be given  by the Berry phase associated with a domain wall Dirac operator when the adiabatic approximation is valid.
We have confirmed this conjecture for a special case in two dimensions.

The Berry phase counts the number of level crossings through Euclidean time development.
This resembles the original APS index theorem, which states that the APS index is counted by the level crossings of massless Dirac operators.
Also, the Berry phase is spatially divided into the bulk and the boundary contributions if the bulk size is sufficiently large.
Each of them gives the bulk integral of Chern-character and the eta-invariant.
We remark that the eta-invariant arises from an integral around the domain walls rather than the exact points where the mass gap closes.
This implies that we do not necessarily have to make domain walls like step functions.
(Indeed, the slope-like mass in \cite{Kanno:2021bze} works as a domain wall.)


It is interesting to generalize our derivation to other higher spatial dimensions and to seek mathematically rigorous formulations.
The new perspective that the APS index is Berry phase would give us better understandings of the non-perturbative aspects of the eta-invariant.
Also, the simplicity of our derivation would enable us to generalize the domain wall APS index theorem to other exotic systems.
For example, the topological nature of non-hermitian systems
would be described by ``Berry phase'' of non-hermitian Dirac operators.
These questions are left for future works.

\subsection*{Acknowledgment}

The authors would like to thank Hidenori Fukaya and Satoshi Yamaguchi for valuable discussions and for making an opportunity to discuss with members of Osaka University Particle Physics Theory Group, Syoto Aoki, Naoki Kawai, Masataka Koide, Yoshiyuki Matsuki, and Yuta Nagoya.
The authors are grateful to Junichi Haruna, Yui Hayashi, Hayato Kanno, and Shuhei Oyama for valuable discussions and daily conversations.
%
The authors appreciate useful discussions during the YITP workshop, ``Topological Phase and Quantum Anomaly 2021'' (YITP-T-21-03), organized by Kantaro Ohmori, Yasunori Lee, Ken Shiozaki, and Yuya Tanizaki.
%
This work was supported in part by
the Sasakawa Scientific Research Grant from The Japan Science Society (T.Y.)
and the Japanese Grant-in-Aid for Scientific Research (No18K03620).

\appendix

\section{Conventions}
\label{sec:conventions}

The Gamma matrices are all chosen to be hermitian as
\begin{align}
	&\{\gamma^{\mu},\gamma^{\nu}\} = 2\delta^{\mu\nu}, \quad
	(\gamma^{\mu})^{\dag} = \gamma^{\nu}, \\
	&\{\bar{\gamma},\gamma^{\mu}\} = 0,
	\quad (\bar{\gamma})^{\dag} = \bar{\gamma}.
	%
\end{align}
The Dirac operator is a Hermitian operator on $\mathcal{M}$ such that
\begin{align}
	D
	= \bar{\gamma}\fs{D}
	= \bar{\gamma}\gamma^{\mu}(\ptl_{\mu}+iA_{\mu}).
\end{align}
Under the chiral representation, we write it as
\begin{align}
	\bar{\gamma} = \mqty(1&0\\0&-1), \quad
	D = \mqty(0&D_{-}\\D_{+}&0).
\end{align}

\section{Eta-invariant in two-dimensional case}
\label{sec:eta_2dim}

See also, for example, App.~A of \cite{Fukaya:2017tsq} for a two-dimensional example of the APS index theorem.
Let us denote the dimensionless eigenvalues of the boundary Dirac operator $i\fs{\nabla}$ by
\begin{align}
	\lambda_n = n+a.
\end{align}
We assume that the holonomy is non-integer.
The eta-invariant is formally defined by
\begin{align}
	\label{eq:eta_form}
	\eta
	= \sum_n \sgn\lambda_n
	= \sum_{n+a>0}1 + \sum_{n+a<0}(-1).
\end{align}
A convenient regularization is the zeta-regularization
\begin{align}
	\eta
	= \sum_n \frac{\sgn\lambda_n}{\abs{\lambda_n}^s} \quad
	(s\rightarrow0).
\end{align}
Indeed, it is evaluated by the Hurwitz zeta function as
\begin{align}
	\eta(s)
	&= \sum_{n+a>0}\frac{1}{(n+a)^s} + \sum_{n+a<0}\frac{-1}{(n+a)^s} \notag\\
	&= \sum_{n=0}^{\infty}\frac{1}{(n+a-[a])^s}
			+\sum_{n=0}^{\infty}\frac{-1}{(n+1-a+[a])^s} \notag\\
	&= \zeta(s,a-[a]) - \zeta(s,1-a+[a]).
\end{align}
Noting that
\begin{align}
	\zeta(0,a) = \frac{1}{2}-a,
\end{align}
the eta-invariant is expressed as
\begin{align}
	\label{eq:eta_reg}
	\frac{\eta}{2}
	= \frac{1}{2}-a+[a].
\end{align}

\section{Landau-Zener effect}
\label{sec:lz}

See also \cite{10011873546,Zener}.
Let us consider a two-state system whose snapshot Hamiltonian is given by
\begin{align}
	h(\tau)
	= \mqty(\varepsilon_1&\varepsilon_{12} \\ \varepsilon_{12}&-\varepsilon_1), \quad
	2\varepsilon_1=\alpha\tau,
\end{align}
where $\alpha,\varepsilon_{12}>0$.
This system has positive/negative energy eigenstates such that
\begin{align}
	h(\tau)\Psi^{\pm}(\tau) = \pm\sqrt{\varepsilon_1^2+\varepsilon_{12}^2}\Psi^{\pm}(\tau).
\end{align}
The energy gap becomes small at $\tau=0$.
Then, the positive/negative energy eigenstates may mix, violating the adiabatic approximation.
The probability of non-adiabatic transition through the dangerous time $\tau=0$ is given by
\begin{align}
	\label{eq:lz_prob}
	P = e^{-2\pi\gamma}, \quad
	\gamma = \frac{\varepsilon_{12}^2}{\alpha}.
\end{align}
Thus, the adiabatic approximation is valid as long as
\begin{align}
	\gamma \gg 1.
\end{align}

\bibliographystyle{utphys}
\bibliography{aps-sqm_ref.bib}

\end{document}